\documentclass{mem}
\usepackage{natbib}\usepackage{txfonts}\usepackage{balance}
\usepackage{graphicx}
\usepackage[a4paper,breaklinks,dvipdfm]{hyperref}
\idline{1}{1}

\def\apjl{ApJ}%
\def\apjs{ApJS}%
\def\ao{Appl.~Opt.}%
\def\aap{A\&A}%
\DeclareSymbolFont{UPM}{U}{eur}{m}{n}
\SetSymbolFont{UPM}{bold}{U}{eur}{b}{n}
\DeclareSymbolFont{AMSa}{U}{msa}{m}{n}
\DeclareMathSymbol{\upi}{0}{UPM}{"19}
\DeclareMathSymbol{\umu}{0}{UPM}{"16}
\DeclareMathSymbol{\upartial}{0}{UPM}{"40}

\SetSymbolFont{symbols}{bold}{OMS}{cmsy}{b}{n}
\DeclareSymbolFont{bmisymbols}{OML}{cmm}{b}{it}
\DeclareMathSymbol{\balpha}{0}{bmisymbols}{"0B}
\DeclareMathSymbol{\bbeta}{0}{bmisymbols}{"0C}
\DeclareMathSymbol{\bgamma}{0}{bmisymbols}{"0D}
\DeclareMathSymbol{\bdelta}{0}{bmisymbols}{"0E}
\DeclareMathSymbol{\bepsilon}{0}{bmisymbols}{"0F}
\DeclareMathSymbol{\bzeta}{0}{bmisymbols}{"10}
\DeclareMathSymbol{\boldeta}{0}{bmisymbols}{"11}
\DeclareMathSymbol{\btheta}{0}{bmisymbols}{"12}
\DeclareMathSymbol{\biota}{0}{bmisymbols}{"13}
\DeclareMathSymbol{\bkappa}{0}{bmisymbols}{"14}
\DeclareMathSymbol{\blambda}{0}{bmisymbols}{"15}
\DeclareMathSymbol{\bmu}{0}{bmisymbols}{"16}
\DeclareMathSymbol{\bnu}{0}{bmisymbols}{"17}
\DeclareMathSymbol{\bxi}{0}{bmisymbols}{"18}
\DeclareMathSymbol{\bpi}{0}{bmisymbols}{"19}
\DeclareMathSymbol{\brho}{0}{bmisymbols}{"1A}
\DeclareMathSymbol{\bsigma}{0}{bmisymbols}{"1B}
\DeclareMathSymbol{\btau}{0}{bmisymbols}{"1C}
\DeclareMathSymbol{\bupsilon}{0}{bmisymbols}{"1D}
\DeclareMathSymbol{\bphi}{0}{bmisymbols}{"1E}
\DeclareMathSymbol{\bchi}{0}{bmisymbols}{"1F}
\DeclareMathSymbol{\bpsi}{0}{bmisymbols}{"20}
\DeclareMathSymbol{\bomega}{0}{bmisymbols}{"21}
\DeclareMathSymbol{\bvarepsilon}{0}{bmisymbols}{"22}
\DeclareMathSymbol{\bvartheta}{0}{bmisymbols}{"23}
\DeclareMathSymbol{\bvarpi}{0}{bmisymbols}{"24}
\DeclareMathSymbol{\bvarrho}{0}{bmisymbols}{"25}
\DeclareMathSymbol{\bvarsigma}{0}{bmisymbols}{"26}
\DeclareMathSymbol{\bvarphi}{0}{bmisymbols}{"27}

\newcommand{\rmn}{\mathrm}

\newcommand{\vel}{\upsilon}
\newcommand{\vvel}{\bupsilon}

\newcommand{\eps}{\varepsilon}

\begin{document}

\title{
Cosmic ray transport in galaxy clusters: implications for radio halos and gamma-rays}

\titlerunning{Cosmic ray transport in galaxy clusters}

\author{
Christoph Pfrommer\inst{1} \and
Torsten En{\ss}lin\inst{2} \and  
Francesco Miniati\inst{3} \and 
Kandaswamy Subramanian\inst{4}}

\authorrunning{Pfrommer et al.}

\institute{
Heidelberg Institute for Theoretical Studies, Schloss-Wolfsbrunnenweg 35, D-69118 Heidelberg, Germany,
\email{christoph.pfrommer@h-its.org}
\and
Max-Planck-Institut f\"ur Astrophysik, Karl-Schwarzschild-Str. 1, D-85741 Garching, Germany
\and 
ETH Zurich Institute of Astronomy, Physics Department, HIT J 12.2. Wolfgang-Pauli-Strasse 27. CH-8093 Zurich, Switzerland
\and 
Inter-University Centre for Astronomy \& Astrophysics, Post Bag 4, Ganeshkhind, Pune 411 007, India
}

\offprints{C. Pfrommer}
 

\abstract{Observations of giant radio halos provide unambiguous evidence for the
  existence of cosmic ray (CR) electrons and magnetic fields in galaxy
  clusters. The physical mechanism generating radio halos is still heavily
  debated.  We critically discuss the proposed models for the radio halo
  emission and highlight the weaknesses underlying each explanation. We present
  an idea how the interplay of CR propagation and turbulent advection selects a
  bimodal spatial CR distribution that is characteristic for the dynamical state
  of a cluster. As a result, strongly turbulent, merging clusters should have a
  more centrally concentrated CR energy density profile with respect to relaxed
  ones with very subsonic turbulence. This translates into a bimodality of the
  expected diffuse radio and gamma ray emission of clusters.  Thus, the observed
  bimodality of cluster radio halos appears to be a natural consequence of the
  interplay of CR transport processes, independent of the model of radio halo
  formation, be it hadronic interactions of CR protons or re-acceleration of
  low-energy CR electrons.  }

\maketitle{}

\section{Introduction}

Relativistic particle populations, cosmic rays (CRs), are expected to permeate
the intra-cluster medium (ICM). Cosmic ray electrons (CRes) are directly visible
in many galaxy clusters via their radio synchrotron emission, forming the
so-called cluster radio halos. Several CRe injection sites can also be
identified via the same synchrotron radiation mechanism: shock waves from
structure formation, active galactic nuclei (AGN), and winds or gas stripping
from cluster galaxies. All these should also be injection sites for CR protons
(CRps) and heavier relativistic nuclei. Due to their higher masses with respect
to the electrons, protons and nuclei are accelerated more efficiently. In our
own Galaxy, the ratio of the spectral energy flux of CRps to CRes between 1\ldots
10 GeV is about one hundred. Similar ratios are also expected at least for the
injection from galaxies and structure formation shock waves for the same
kinematic reasons.\footnote{For an extended list of references and a more
  detailed discussion, see \citet{2011A&A...527A..99E} which our proceeding
  closely follows.}

Cluster CRps should have accumulated over cosmic timescales since the bulk of
them is unable to leave through the persistent infall of matter onto the cluster
and due to the long CRps' radiative lifetimes in the ICM of the order of an
Hubble time throughout the entire ICM. CRes suffer much more severe energy losses
via synchrotron and inverse Compton emission at GeV energies, and Bremsstrahlung
and Coulomb losses below 100 MeV. CRes with an energy of $\sim 10~$GeV emit GHz
synchrotron waves in $\mu$G-strength magnetic fields. Since the associated
inverse Compton and synchrotron cooling time is $\tau_\rmn{IC,syn}\sim 10^8$~yr,
these CRes must have been recently injected or re-accelerated.

\section{Observations of radio halos}

\begin{figure*}
 \centering
 \includegraphics[bb=135 253 478 534, width=0.49\textwidth]{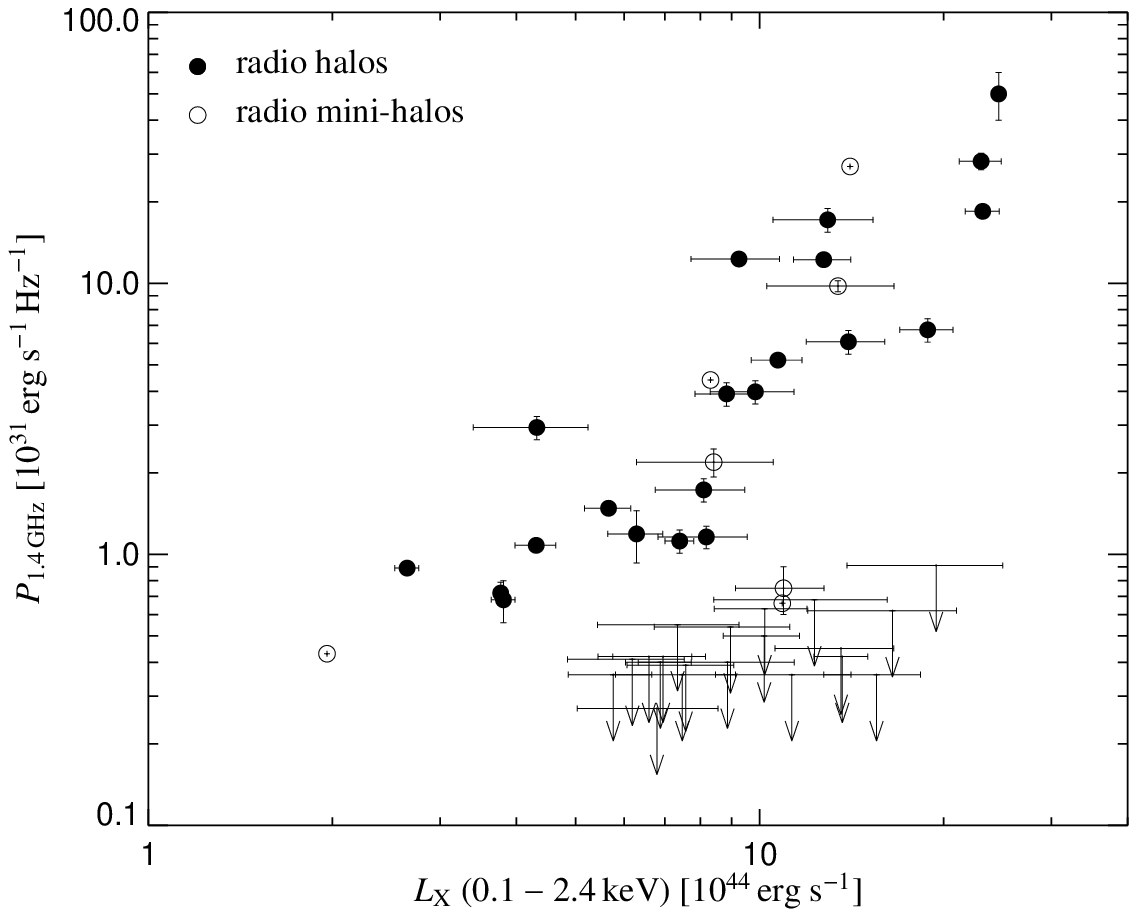}
\includegraphics[bb=135 253 478 534, width=0.49\textwidth]{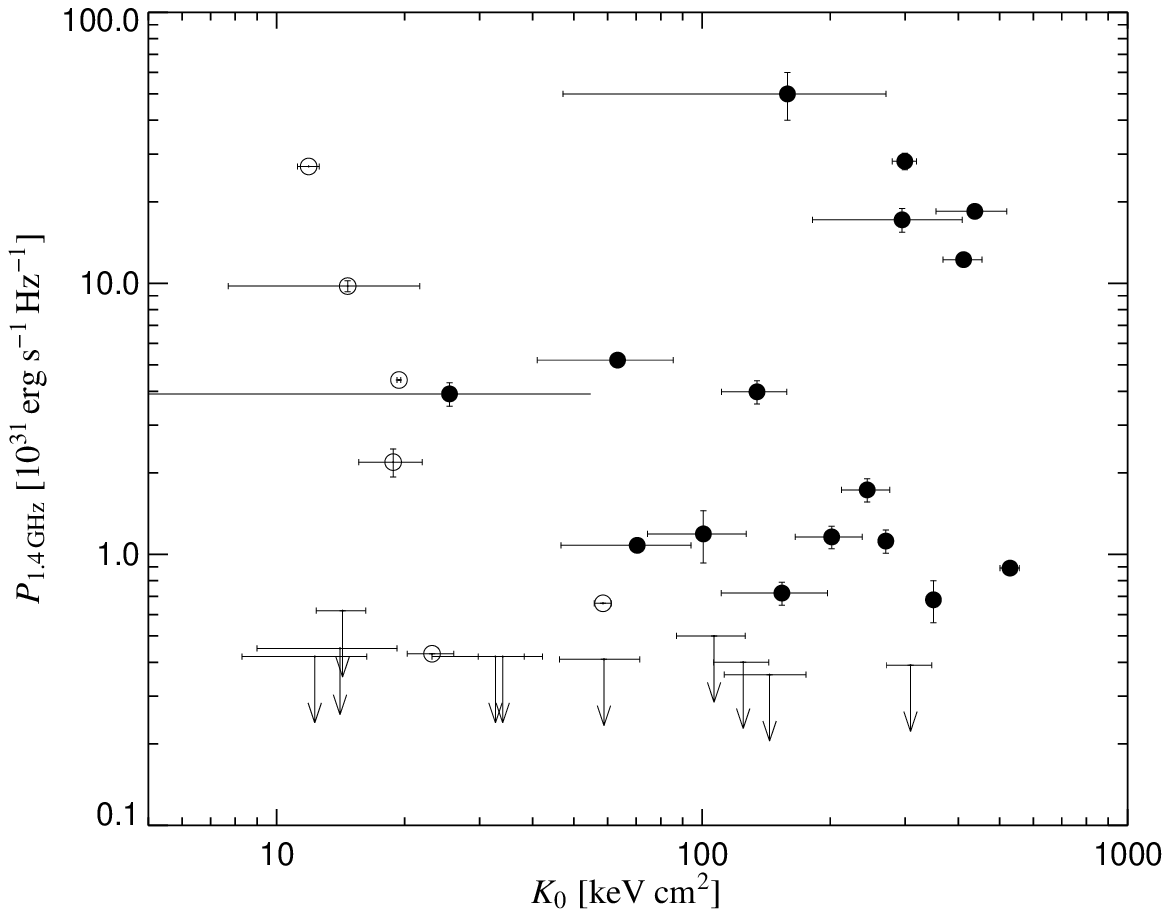}
\caption{Correlation of radio halo luminosities with cluster properties
  \citep{2011A&A...527A..99E}.  \emph{Left:} Radio halo luminosity vs X-ray
  luminosity.  \emph{Right:} Radio halo luminosity vs central entropy
  indicator $K_0$ for the subsample of clusters for which high resolution
  Chandra data are available.  }
 \label{fig:radiohalo}
\end{figure*}

Cluster radio halos are our primary evidence for the existence of CRs in galaxy
clusters. They are spatially extended regions of diffuse radio emission, which
have regular morphologies (resembling the morphology of the X-ray emitting
thermal ICM plasma). Their radio synchrotron emission is unpolarised, due to the
contribution of various magnetic field orientations along the line of sight, and
Faraday rotation de-polarisation.

Cluster radio halos come in two sizes: cluster wide and therefore giant radio
halos and radio mini-halos. The former are predominantly found in clusters
showing merger activities whereas the latter are found in very relaxed clusters
which developed a cool core that harbors the mini-halo. The radio (mini-)halo
luminosity correlates with the X-ray emissivity of the cluster (see
Fig.~\ref{fig:radiohalo}).  A large fraction of clusters do not exhibit
significant radio halo emission, and only upper limits to their
synchrotron flux are known. About half of the radio deficient clusters, for
which we have Chandra data, show clear evidence for some level of cool core
structure ($K_0 \lesssim 40\,\rmn{keV\,cm^2}$) as can be seen in
Fig.~\ref{fig:radiohalo}. This could either imply that these clusters are in the
intermediate state between having giant radio halos because of merging activity
and having mini halos due to strongly developed cool cores. On the other hand
there could be two populations of clusters -- cool cores and non-cool cores --
and the corresponding radio luminosity responds sensitively to the level of
injected turbulence by either AGN or cluster mergers, respectively.

\subsection{Hadronic models}
In the hadronic model the accumulated CRps continuously inject radio
emitting CRes into the ICM due to well known hadronic process $p_\mathrm{CR} + p
\rightarrow \pi^{\pm} + \ldots \rightarrow e^{\pm} +
\nu_\rmn{e}/\bar{\nu}_\rmn{e} + \nu_\mu+\bar{\nu}_\mu + \dots$  The hadronic
model has \emph{advantages}:
\begin{itemize}
\item All required ingredients are available: ample sources of CRps (structure
  formation shocks, AGN, galactic winds), gas protons as targets, magnetic
  fields.
\item Smooth and regular morphology of halos are a consequence of the long
  lifetime of CRps which implies a volume filling cluster distribution.
\item Using analytical arguments and hydrodynamical simulations, the predicted
  luminosities, scalings ($L_\nu-L_X$), and morphologies match observations
  without tuning \citep{2001ApJ...562..233M, 2008MNRAS.385.1211P,
    2009JCAP...09..024K}.
\item The model predicts power-law spectra as observed.
\end{itemize}
There are also \emph{issues} with the hadronic model:
\begin{itemize}
\item About two thirds of the most X-ray luminous clusters do not exhibit radio
  halos, whereas the hadronic model seems to suggests that all clusters exhibit
  halos [will be addressed in the following].
\item The model does not explain all reported spectral features (curvature,
  spectral steepening) [will be addressed in the following].
\end{itemize}
The hadronic model makes a testable prediction: the radio halo emission should
always be accompanied by weak diffuse gamma-ray emission, due to the hadronic
production of neutral pions and their decay into gamma-rays, $p_\mathrm{CR} + p
\rightarrow \pi^{0} + \ldots\rightarrow 2\, \gamma+ \ldots$ The current upper
limits on diffuse gamma-ray flux from cluster of galaxies by the Fermi
collaboration \citep{2010ApJ...717L..71A} are still well above the predictions
of expected fluxes, even for the most optimistic assumptions about the CR
acceleration efficiency \citep{2010MNRAS.409..449P}. They are far off the
minimal gamma-ray flux expected in the limit of strong magnetic field strength
\citep[$\gg 3 \mu$G;][]{2008MNRAS.385.1242P, 2010ApJ...710..634A}.

\begin{figure*}
 \includegraphics[bb=0 0 653 181,width=\textwidth]{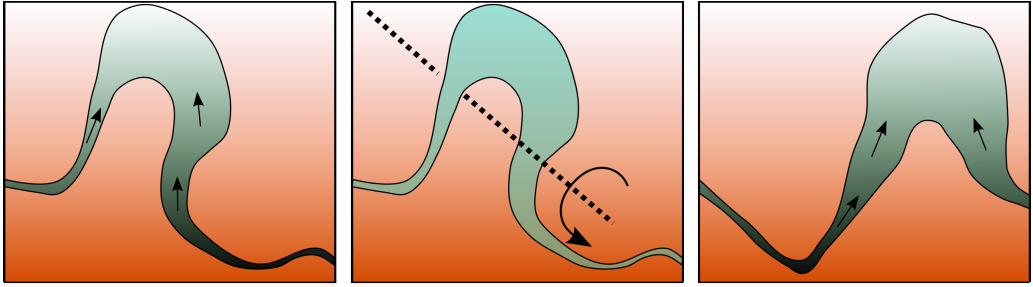}
 \caption{Sketch of the interplay of CR streaming and turbulent advection for a
   single flux tube in a stratified atmosphere with gravity pointing
   downwards. \emph{Left:} The dense CRs at the center stream along the tube
   towards the CR depleted regions at larger atmospheric
   height. \emph{Middle:} CR streaming stops as soon as a homogeneous CR space
   density is achieved. A turbulent eddy (represented by its angular momentum
   axis) starts to turn the magnetic structure upside down. \emph{Right:} The
   former outer parts of the flux tubes are compressed at the center, and harbor
   now an overdense CR population, whereas the former inner parts are expanded
   at larger atmospheric scale height and therefore have now an underdense CR
   population. Again CR streaming sets in.}
 \label{fig:sketch1}
\end{figure*}

\subsection{Re-acceleration models}
In re-acceleration models, a pre-existing CRe population at lower energies of
about 0.1-10 GeV gets re-accelerated into the radio emitting regime of about 10
GeV by plasma waves.  These are generated by the turbulence during and after a
cluster merger event.  Some level of re-acceleration has to happen most of the
time or frequently enough in order to prevent the CRe population in the cluster
center from loosing its energy completely due to Coulomb losses on a timescale
of about 1 Gyr. The \emph{advantages} of the re-acceleration model are:
\begin{itemize}
\item All required ingredients are available: radio galaxies and relics to
  inject CRes and plasma waves to re-accelerate them.
\item The bimodality of radio halo luminosities is explained by
  the presence and decay of the re-accelerating turbulence in merging and
  relaxed clusters, respectively \citep{2009A&A...507..661B}.
\item Reported complex radio spectra in some clusters emerge naturally by
  interplay of of acceleration and cooling.
\end{itemize}
The \emph{issues} with the re-acceleration model are:
\begin{itemize}
\item Fermi II acceleration is inefficient and scales with $\vel_\rmn{wave}^2
  /c^2 \ll 1$; efficiency in current models is fitted to explain data and not
  derived from first principles.
\item Current models neglect advective energy losses by waves that propagate
  outwards and dissipate in the outer regions.
\item Intermittency of turbulence might be difficult to reconcile with the
  observed regularity of radio halos.
\item Observed power-law spectra require fine tuning.
\item CRes cool rapidly in the central regions on timescale of 1 Gyr (for
  $n_\rmn{e}=3\times10^{-3}\,\rmn{cm}^{-3}$) [will be addressed in the
  following].
\end{itemize}
A testable prediction with upcoming sensitive radio telescope arrays is that
low X-ray luminous clusters should not exhibit radio halos
\citep{2008A&A...480..687C}.

\section{Cosmic ray transport}

\begin{figure*}
 \includegraphics[bb=25 180 550 600,width=0.49\textwidth]{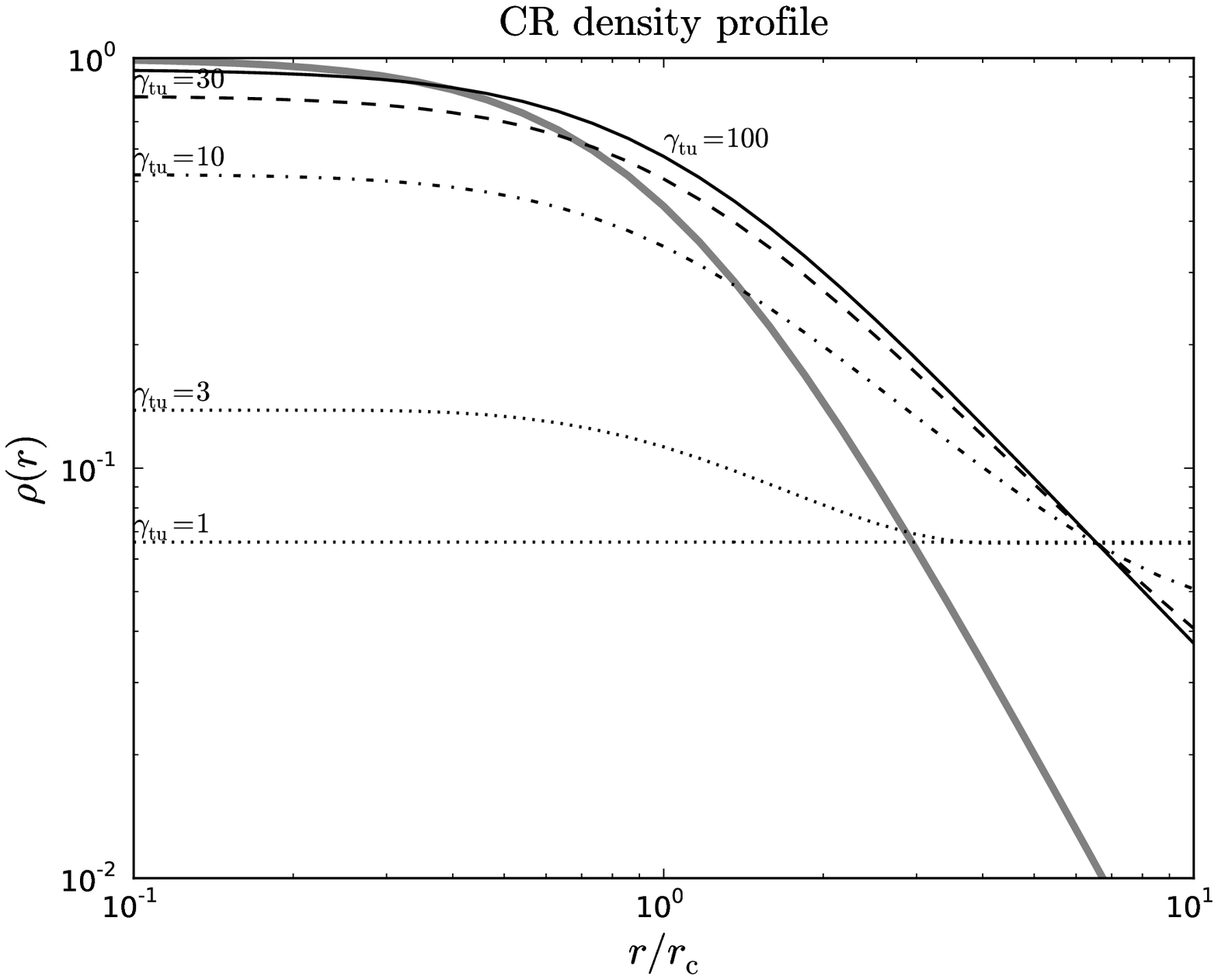}
 \includegraphics[bb=25 180 550 600,width=0.49\textwidth]{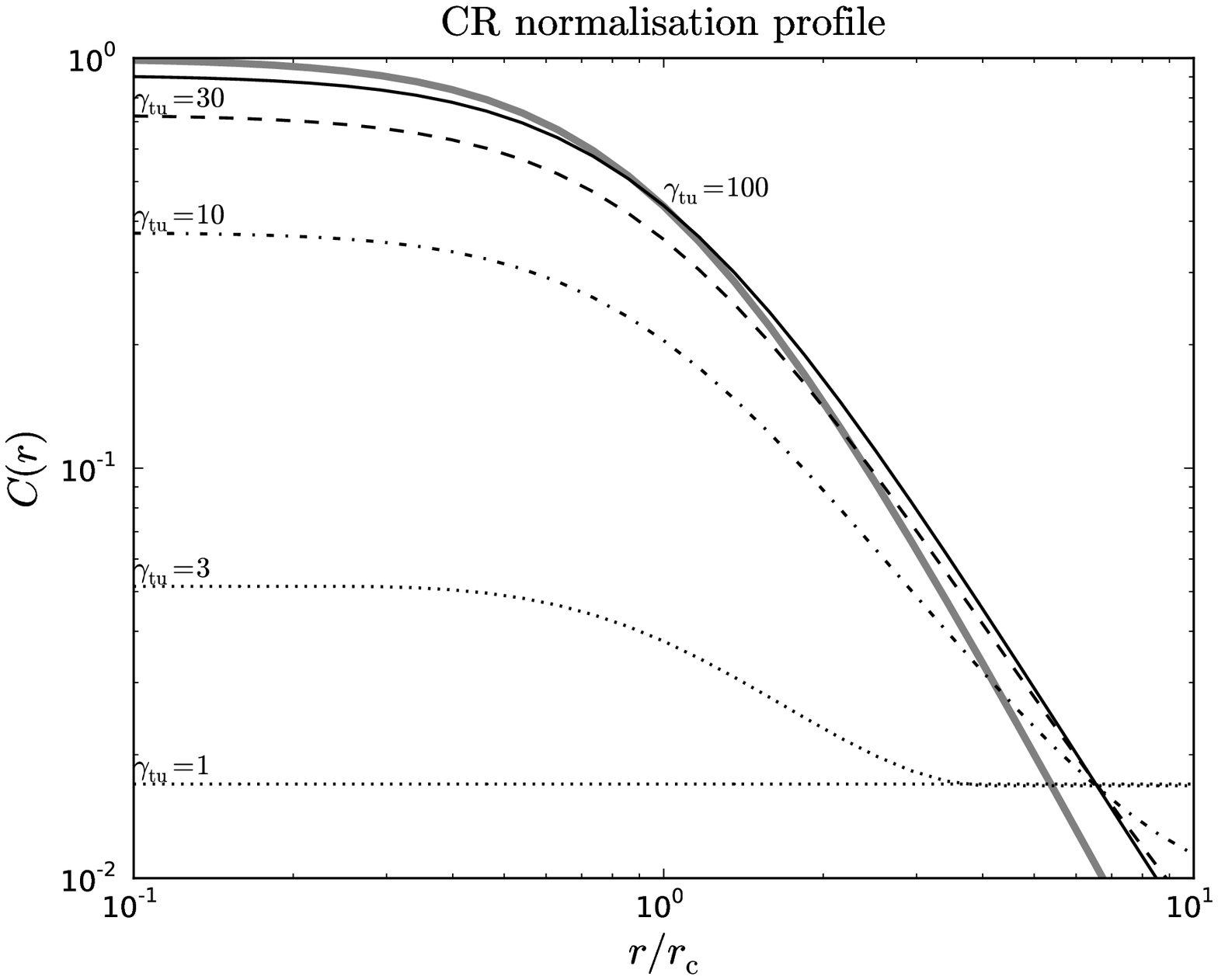}
 \caption{\emph{Left:} CR density profiles for $\gamma_\rmn{tu} = 1, 3,\,10, \,
   30,$ and $ 100 $ (from bottom to top at small radii) including the same
   number of CRs each. Profiles are normalised to
   $\varrho(0)|_{\gamma_\rmn{tu}=\infty}$. Also the more narrow gas density
   profile is shown (thick grey line). \emph{Right:} CR normalisation profiles
   for the same parameters and the gas density profile for typical cluster
   conditions.}
 \label{fig:density}
\end{figure*}

\subsection{Confined cosmic rays}

To begin, we consider an isolated magnetic flux tube with CRs confined to it to
illustrate the interplay of advection and streaming with a basic picture. This
represents the limiting case of \emph{confined CRs} and will be generalised in
the next section. Imagine a magnetic flux tube frozen into the plasma which is
distributed in a stratified pressure atmosphere of a cluster as shown in
Fig. \ref{fig:sketch1} on the left. Any central concentration of CRs will escape
due to CR streaming on a timescale of $\tau_\mathrm{st} = L_B /
\vel_\mathrm{st}$, where $L_B$ is the magnetic bending scale and
$\vel_\mathrm{st}$ the CR streaming velocity along the magnetic field which is
of order the sound speed in the cluster plasma.\footnote{In a low-$\beta$
  plasma, the CR streaming velocity is linked to the Alfv\'en velocity, which
  exceeds the sound speed there.  However, this can obviously not be true in a
  high-$\beta$ plasma.  It would imply that in the limit of vanishing magnetic
  field strength the CRs get completely immobile due to the vanishing Alfv\'en
  speed.  However, for disappearing magnetic fields, the coupling of CRs to the
  plasma gets weaker and therefore the CRs should stream faster.  Thus, there
  must be a characteristic velocity, below which the Alfv\'en velocity is not
  limiting the streaming velocity any more.  Plasma physical arguments indicate
  that this is roughly the sound speed \citep{2001ApJ...553..198F,
    2011A&A...527A..99E}.}  This leads to a homogeneous CR distribution within
the flux tube (Fig. \ref{fig:sketch1}, middle). Turbulence turns the magnetic
structure upside down on half an eddy turnover time. If this is comparable to,
or less than, the CR escape time,
\begin{equation}
 \frac{\tau_\mathrm{st}}{\tau_\mathrm{tu}} 
\equiv \gamma_\rmn{tu} \sim \mathcal{O}(1),
\end{equation}
a good fraction of the CRs from larger radii will be compressed towards the
center, from where they again start streaming to larger radii.  \emph{The
  transonic turbulence is therefore able to maintain a centrally enhanced CR
  density by pumping expanded CR populations downwards.} As soon as the
turbulent velocities become significantly subsonic, this pumping becomes
inefficient, since the streaming will be faster than the advection. At this
point a nearly constant volume density of CRs establishes within a closed flux
tube, meaning that most CRs are residing at larger cluster radii. Depending on
the level of turbulence, we obtain either a CR distribution that is peaked
towards the center or a homogeneous CR distribution.

\subsection{Mobile cosmic rays}

In reality, CR diffusion perpendicular to the mean magnetic field enables CRs to
change between magnetic flux tubes and thereby find paths to more peripheral
regions. The accessible distance is determined by the level of turbulent
pumping, magnetic topology, and available time to stream. In principle, CRs can
even reach the outskirts of galaxy clusters, where the infall of matter onto the
cluster behind the accretion shocks prevents further escape which motivates our
term \emph{mobile CRs}.

We assume a power-law CR spectrum,
\begin{equation} 
\label{eq:CRspec}
f(\vec{r},p,t) = C(\vec{r},t)\, p^{-\alpha},
\end{equation}
where $\alpha \approx 2.1-2.5$ is the spectral index and $C(\vec{r},t)$ the
spectral normalisation constant.  First, we derive the equilibrium profile that
CRs attain if turbulent advection dominates the CR transport (and CR streaming
is negligible). To this end, we assume that (i) the cluster is characterised by
a mean pressure profile and (ii) that CR propagation operates on small scales,
permitting CR exchange between nearby gas volume elements, but not on large
scales.  Whenever two volume elements come close, CRs can be exchanged which
establishes a constant CR population in any given radial shell. During radial
advective transport from radius $r$ to $r'$, the ICM gas with the entrained CRs
is compressed or expanded by a factor $X(r \rightarrow r') =
(P(r')/P(r))^{1/\gamma}$, where $P(r)$ is the pressure profile and $\gamma=5/3$.
The CR rest-mass density $\varrho(\vec{r})= m\,\int dp\, f(\vec{r},p)$ thus
establishes -- under the influence of advection alone -- a profile according to
\begin{eqnarray}
\label{eq:varrhowithP}
 \varrho(r) &=& \varrho_0\, \left( \frac{P(r)}{P_0} \right)^{\frac{1}{\gamma}}=   \varrho_0 \, \eta(r),
\end{eqnarray}
where $\eta(r) = (P(r)/P_0)^{1/\gamma}$ is the advective CR target profile.  

The CR continuity equation for $\varrho$ in the absence of sources and sinks can
be written as
\begin{equation}
 \frac{\partial  \varrho}{\partial t} + \vec{\nabla}\cdot(\vvel\, \varrho) = 0,
\end{equation}
with $\vvel = \vvel_\mathrm{ad} + \vvel_\mathrm{di} + \vvel_\mathrm{st}$ the CR
transport velocity, which is composed of the advective ($\vvel_\mathrm{ad}$),
diffusive ($\vvel_\mathrm{di}$), and steaming ($\vvel_\mathrm{st}$) transport
velocities. These are defined by
\begin{eqnarray}
 \vvel_\mathrm{st} &=&  - \vel_\mathrm{st}\, \frac{\vec{\nabla} \, \varrho}{|\vec{\nabla} \, \varrho|}, \nonumber\\
 \vvel_\mathrm{di} &=&  - {\kappa}_\mathrm{di}\, \, \frac{1}{\varrho} \, \vec{\nabla} \varrho
                   =   - \kappa_\mathrm{di} \, \vec{\nabla} \ln(\varrho),  \\
 \vvel_\mathrm{ad} &=&  - \kappa_\mathrm{tu} \, \frac{\eta}{\varrho} \, \vec{\nabla} \frac{\varrho}{\eta}
                   =   - \kappa_\mathrm{tu} \, \vec{\nabla} \ln \left( \frac{\varrho}{\eta}\right), \nonumber
\end{eqnarray}
where ${\kappa}_\mathrm{di}$ is the macroscopically averaged CR diffusion
coefficient and the passive, advective transport via turbulence can be described
by an additional diffusion process with diffusion coefficient
$\kappa_\mathrm{tu} = L_\mathrm{tu}\, \vel_\mathrm{tu}/3$. We note that the
appearance of the target density profile $\eta(\vec{r})$ in the gradient for
$\vvel_\mathrm{ad}$ ensures that any deviation of the CR distribution from
target density causes a restoring term towards this equilibrium configuration.
The CR space density becomes stationary for $\vvel = \mathbf{0}$, and this reads
in spherical symmetry with radially outstreaming CRs
\begin{equation}
\label{eq:transpSpherical}
  {\vel}_\mathrm{st} =  \kappa_\mathrm{tu} \, \frac{\partial}{\partial r} 
  \ln \left( \frac{\varrho}{\eta}\right)  + 
  \kappa_\mathrm{di}\,  \frac{\partial}{\partial r}  \ln(\varrho).
\end{equation} 
\citet{2011A&A...527A..99E} provide an analytical solution of this equation which is
shown in Fig.~\ref{fig:density} for different values of $\gamma_\rmn{tu}$.

\section{Radio and gamma-ray bimodality}

\begin{figure*}
 \includegraphics[bb=25 180 550 600,width=0.49\textwidth]{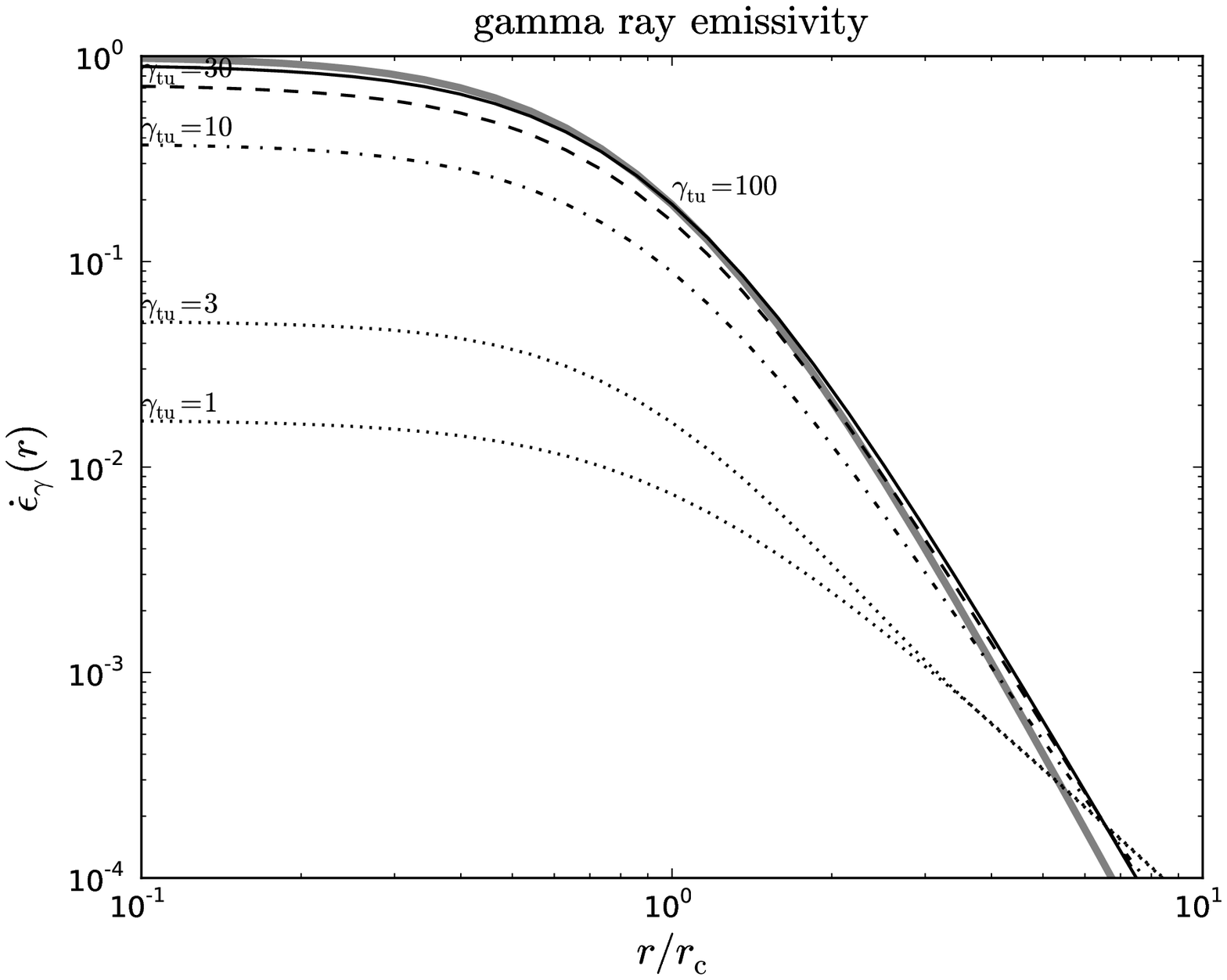}
\includegraphics[bb=25 180 550 600,width=0.49\textwidth]{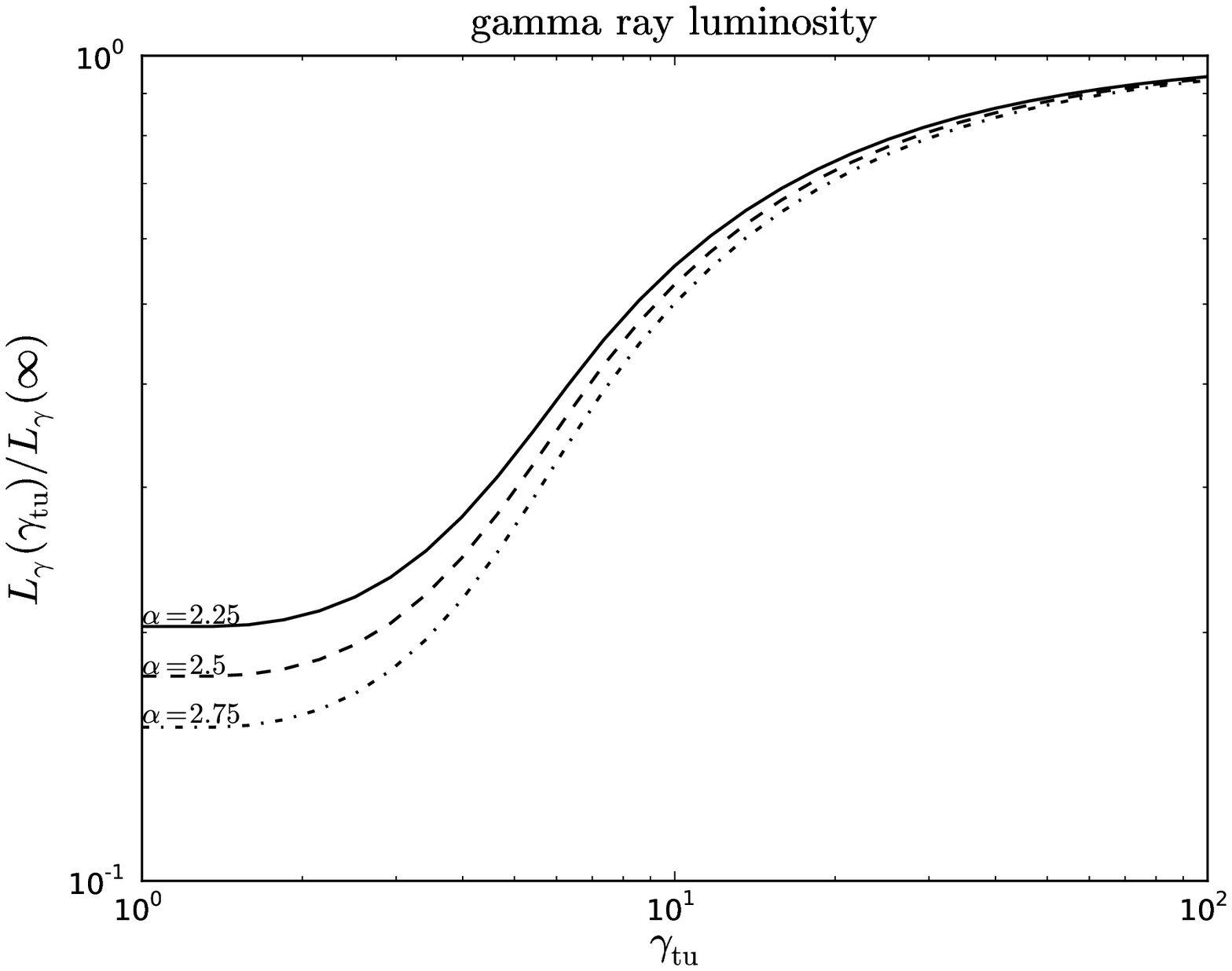}
\caption{\emph{Left:} Gamma-ray emissivity profiles for the CR distributions
  in Fig. \ref{fig:density} and X-ray emissivity profile of the ICM in
    grey. Emissivities are normalised to the central emissivity of a cluster
    with $\gamma_\rmn{tu}=\infty$.  \emph{Right:} Total gamma-ray flux due
  to hadronic CRp interactions with the ICM nucleons as a function of
  $\gamma_\rmn{tu}={\tau}_\rmn{st}/\tau_\rmn{ad}$ and for $\alpha = 2.25$,
  $2.5$, and $2.75$ (solid, dashed, and dashed-dotted lines, respectively).
  Normalised to $L_\gamma$ for $\gamma_\rmn{tu}=\infty$.}
 \label{fig:gamma}
\end{figure*}

\begin{figure*}
 \includegraphics[bb=25 180 550 600,width=0.49\textwidth]{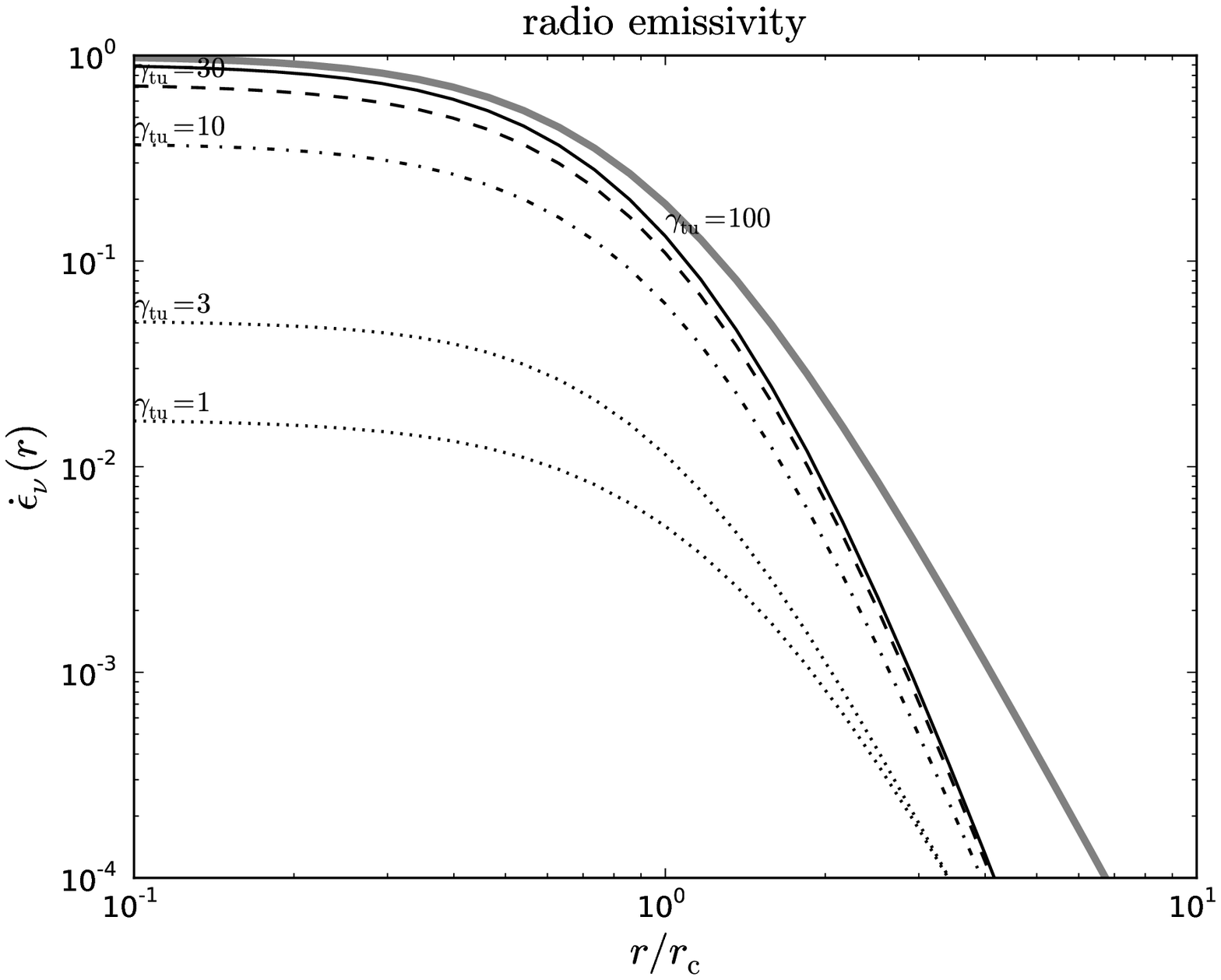}
 \includegraphics[bb=25 180 550 600,width=0.49\textwidth]{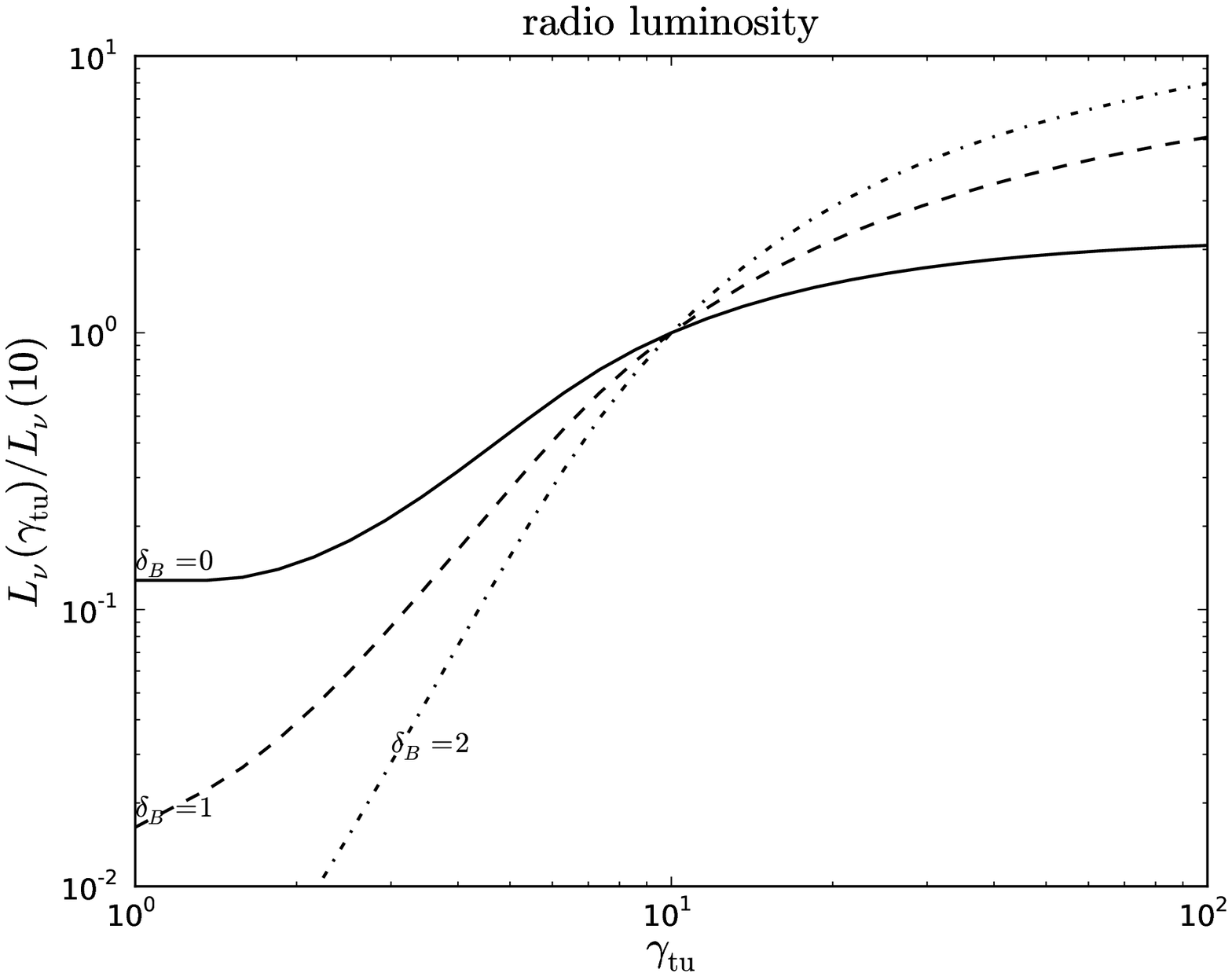}
 \caption{\emph{Left:} Radio emissivity profiles for the cluster shown in
   Fig. \ref{fig:density} assuming the same magnetic field profiles with $B_0 =
   6\,\mu\rmn{G}$ and $\delta_B = 0$. Emissivities are normalised to the central
   radio emissivity of a cluster with $\gamma_\rmn{tu}=\infty$. The X-ray
   profile is shown in grey. \emph{Right:} Total radio flux due to hadronic CRp
   interactions with the ICM nucleons as a function of
   $\gamma_\rmn{tu}={\tau}_\rmn{st}/\tau_\rmn{ad}$ and for different
   dependencies of the magnetic energy density on the turbulence level,
   $\eps_B\propto n(r) \gamma_\rmn{tu}^{\delta_B}$ and parametrised by $\delta_B
   = 0$, $1$, and $2$ (solid, dashed, and dashed-dotted lines, respectively).
   Normalised to $L_\nu$ for $\gamma_\rmn{tu}=10$. (Central field strength $B_0
   = 6\,\mu\rmn{G}$, $\alpha=2.5$).}
 \label{fig:radio}
\end{figure*}

The gamma-ray emissivity and luminosity of a power law CRp spectrum as in
Eqn. (\ref{eq:CRspec}) is
\begin{equation}
 \dot{\eps}_\gamma \propto C\, \varrho_\rmn{gas}, \quad\mbox{and~}
  L_\gamma = \int dV\, \dot{\eps}_\gamma.
\end{equation}
The radio luminosity in the hadronic model is 
\begin{equation}
\label{eq:Lnu}
  \dot{\eps}_\nu \propto C \, \varrho_\rmn{gas}\, \frac{\eps_B^{(\alpha+2)/4}}{\eps_B+\eps_\rmn{ph}},
 \quad\mbox{and~}
  L_\nu = \int dV\, \dot{\eps}_\nu.
\end{equation}
As shown in Figs.~\ref{fig:gamma} and \ref{fig:radio}, $L_\gamma$ and $L_\nu$
inherit the strong dependence on the advective-to-streaming-velocity ratio,
$\gamma_\rmn{tu}=\vel_\rmn{tu} /\vel_\rmn{st}$. Thus, a rapid drop in radio
luminosity after the turbulent merger phase by one order of magnitude or more is
actually expected in the \emph{hadronic halo model} on a timescale of 0.1--1
Gyr, depending on magnetic topology and the macroscopic CR streaming speed
\citep{2011A&A...527A..99E}.

\section{Conclusions}

\emph{CR streaming} (and CR diffusion) aims at establishing a spatially flat CR
profile; hence explaining why radio halos are not found in every cluster.
\emph{CR advection} tends to produce centrally enhanced CR profiles.  Thus, CR
advection and streaming are counteracting transport mechanisms.  Whenever the
former dominates, centrally enhanced profiles are established, and whenever
streaming is more important, a flat profile results.

During a cluster merger, advective velocities in galaxy clusters are comparable
to the sound speed and drop when the cluster relaxes after the merger. Plasma
physical arguments suggest that the microscopic CR streaming velocity in
clusters might of the order of the sound speed.  Macroscopically it is
reduced due to magnetic trapping of CRs in flux tubes (which is larger for a
stronger turbulence) and slow cross field diffusion required to escape.

As a result of this, merging clusters should have a much more centrally
concentrated CR population than relaxed ones. This leads naturally to a
\emph{bimodality of their gamma-ray and radio synchrotron emissivities due to hadronic
interactions of CR protons}. Also in the re-acceleration model of cluster radio
halos these transport processes should be essential, since the re-accelerated CR
electron populations in the dense cluster centers is probably too vulnerable to
Coulomb losses, to survive periods without significant
re-acceleration. Transport of the longer living electrons at the cluster
outskirts into the cluster center during cluster merger would circumvent this
problem.

We also expect an energy dependence of the macroscopic CR streaming speed, which
then should lead to a spatial differentiation of the spectral index of the CRp
population and any secondary radio halo emission. Such spectral index variation
in the radio halo should become especially strong during phases of outstreaming
CRps, i.e. when a radio halo dies due to the decay of the cluster turbulence.


\bibliographystyle{aa}
\bibliography{pfrommer}

\begin{thebibliography}{11}
\expandafter\ifx\csname natexlab\endcsname\relax\def\natexlab#1{#1}\fi

\bibitem[{{Ackermann} {et~al.}(2010){Ackermann}, {Ajello}, {Allafort},
  {Baldini}, {Ballet}, {Barbiellini}, {Bastieri}, {Bechtol}, {Bellazzini},
  {Blandford}, {Blasi}, {Bloom}, {Bonamente}, {Borgland}, {Bouvier}, {Brandt},
  {Bregeon}, {Brigida}, {Bruel}, {Buehler}, {Buson}, {Caliandro}, {Cameron},
  {Caraveo}, {Carrigan}, {Casandjian}, {Cavazzuti}, {Cecchi}, {{\c C}elik},
  {Charles}, {Chekhtman}, {Cheung}, {Chiang}, {Ciprini}, {Claus},
  {Cohen-Tanugi}, {Colafrancesco}, {Cominsky}, {Conrad}, {Dermer}, {de Palma},
  {Silva}, {Drell}, {Dubois}, {Dumora}, {Edmonds}, {Farnier}, {Favuzzi},
  {Frailis}, {Fukazawa}, {Funk}, {Fusco}, {Gargano}, {Gasparrini}, {Gehrels},
  {Germani}, {Giglietto}, {Giordano}, {Giroletti}, {Glanzman}, {Godfrey},
  {Grenier}, {Grondin}, {Guiriec}, {Hadasch}, {Harding}, {Hayashida}, {Hays},
  {Horan}, {Hughes}, {Jeltema}, {J{\'o}hannesson}, {Johnson}, {Johnson},
  {Johnson}, {Kamae}, {Katagiri}, {Kataoka}, {Kerr}, {Kn{\"o}dlseder}, {Kuss},
  {Lande}, {Latronico}, {Lee}, {Lemoine-Goumard}, {Longo}, {Loparco}, {Lott},
  {Lovellette}, {Lubrano}, {Madejski}, {Makeev}, {Mazziotta}, {Michelson},
  {Mitthumsiri}, {Mizuno}, {Moiseev}, {Monte}, {Monzani}, {Morselli},
  {Moskalenko}, {Murgia}, {Naumann-Godo}, {Nolan}, {Norris}, {Nuss}, {Ohsugi},
  {Omodei}, {Orlando}, {Ormes}, {Ozaki}, {Paneque}, {Panetta}, {Pepe},
  {Pesce-Rollins}, {Petrosian}, {Pfrommer}, {Piron}, {Porter}, {Profumo},
  {Rain{\`o}}, {Rando}, {Razzano}, {Reimer}, {Reimer}, {Reposeur}, {Ripken},
  {Ritz}, {Rodriguez}, {Romani}, {Roth}, {Sadrozinski}, {Sander}, {Saz
  Parkinson}, {Scargle}, {Sgr{\`o}}, {Siskind}, {Smith}, {Spandre}, {Spinelli},
  {Starck}, {Stawarz}, {Strickman}, {Strong}, {Suson}, {Tajima}, {Takahashi},
  {Takahashi}, {Tanaka}, {Thayer}, {Thayer}, {Tibaldo}, {Tibolla}, {Torres},
  {Tosti}, {Tramacere}, {Uchiyama}, {Usher}, {Vandenbroucke}, {Vasileiou},
  {Vilchez}, {Vitale}, {Waite}, {Wang}, {Winer}, {Wood}, {Yang}, {Ylinen}, \&
  {Ziegler}}]{2010ApJ...717L..71A}
{Ackermann}, M., {Ajello}, M., {Allafort}, A., {et~al.} 2010, \apjl, 717, L71

\bibitem[{{Aleksi{\'c}} {et~al.}(2010){Aleksi{\'c}}, {Antonelli}, {Antoranz},
  {Backes}, {Baixeras}, {Balestra}, {Barrio}, {Bastieri}, {Becerra
  Gonz{\'a}lez}, {Bednarek}, {Berdyugin}, {Berger}, {Bernardini}, {Biland},
  {Bock}, {Bonnoli}, {Bordas}, {Borla Tridon}, {Bosch-Ramon}, {Bose}, {Braun},
  {Bretz}, {Britzger}, {Camara}, {Carmona}, {Carosi}, {Colin}, {Commichau},
  {Contreras}, {Cortina}, {Costado}, {Covino}, {Dazzi}, {De Angelis}, {De Cea
  del Pozo}, {De los Reyes}, {De Lotto}, {De Maria}, {De Sabata}, {Delgado
  Mendez}, {Doert}, {Dom{\'{\i}}nguez}, {Dominis Prester}, {Dorner}, {Doro},
  {Elsaesser}, {Errando}, {Ferenc}, {Fonseca}, {Font}, {Galante},
  {Garc{\'{\i}}a L{\'o}pez}, {Garczarczyk}, {Gaug}, {Godinovic}, {Hadasch},
  {Herrero}, {Hildebrand}, {H{\"o}hne-M{\"o}nch}, {Hose}, {Hrupec}, {Hsu},
  {Jogler}, {Klepser}, {Kr{\"a}henb{\"u}hl}, {Kranich}, {La Barbera}, {Laille},
  {Leonardo}, {Lindfors}, {Lombardi}, {Longo}, {L{\'o}pez}, {Lorenz},
  {Majumdar}, {Maneva}, {Mankuzhiyil}, {Mannheim}, {Maraschi}, {Mariotti},
  {Mart{\'{\i}}nez}, {Mazin}, {Meucci}, {Miranda}, {Mirzoyan}, {Miyamoto},
  {Mold{\'o}n}, {Moles}, {Moralejo}, {Nieto}, {Nilsson}, {Ninkovic}, {Orito},
  {Oya}, {Paiano}, {Paoletti}, {Paredes}, {Partini}, {Pasanen}, {Pascoli},
  {Pauss}, {Pegna}, {Perez-Torres}, {Persic}, {Peruzzo}, {Prada}, {Prandini},
  {Puchades}, {Puljak}, {Reichardt}, {Rhode}, {Rib{\'o}}, {Rico}, {Rissi},
  {R{\"u}gamer}, {Saggion}, {Saito}, {Salvati}, {S{\'a}nchez-Conde},
  {Satalecka}, {Scalzotto}, {Scapin}, {Schultz}, {Schweizer}, {Shayduk},
  {Shore}, {Sierpowska-Bartosik}, {Sillanp{\"a}{\"a}}, {Sitarek}, {Sobczynska},
  {Spanier}, {Spiro}, {Stamerra}, {Steinke}, {Struebig}, {Suric}, {Takalo},
  {Tavecchio}, {Temnikov}, {Terzic}, {Tescaro}, {Teshima}, {Torres}, {Vankov},
  {Wagner}, {Zabalza}, {Zandanel}, {Zanin}, {Zapatero}, {MAGIC Collaboration},
  {Pfrommer}, {Pinzke}, {En{\ss}lin}, {Inoue}, \&
  {Ghisellini}}]{2010ApJ...710..634A}
{Aleksi{\'c}}, J., {Antonelli}, L.~A., {Antoranz}, P., {et~al.} 2010, \apj,
  710, 634

\bibitem[{{Brunetti} {et~al.}(2009){Brunetti}, {Cassano}, {Dolag}, \&
  {Setti}}]{2009A&A...507..661B}
{Brunetti}, G., {Cassano}, R., {Dolag}, K., \& {Setti}, G. 2009, \aap, 507, 661

\bibitem[{{Cassano} {et~al.}(2008){Cassano}, {Brunetti}, {Venturi}, {Setti},
  {Dallacasa}, {Giacintucci}, \& {Bardelli}}]{2008A&A...480..687C}
{Cassano}, R., {Brunetti}, G., {Venturi}, T., {et~al.} 2008, \aap, 480, 687

\bibitem[{{En{\ss}lin} {et~al.}(2011){En{\ss}lin}, {Pfrommer}, {Miniati}, \&
  {Subramanian}}]{2011A&A...527A..99E}
{En{\ss}lin}, T., {Pfrommer}, C., {Miniati}, F., \& {Subramanian}, K. 2011,
  \aap, 527, A99+

\bibitem[{{Felice} \& {Kulsrud}(2001)}]{2001ApJ...553..198F}
{Felice}, G.~M. \& {Kulsrud}, R.~M. 2001, \apj, 553, 198

\bibitem[{{Kushnir} {et~al.}(2009){Kushnir}, {Katz}, \&
  {Waxman}}]{2009JCAP...09..024K}
{Kushnir}, D., {Katz}, B., \& {Waxman}, E. 2009, \jcap, 9, 24

\bibitem[{{Miniati} {et~al.}(2001){Miniati}, {Jones}, {Kang}, \&
  {Ryu}}]{2001ApJ...562..233M}
{Miniati}, F., {Jones}, T.~W., {Kang}, H., \& {Ryu}, D. 2001, \apj, 562, 233

\bibitem[{{Pfrommer}(2008)}]{2008MNRAS.385.1242P}
{Pfrommer}, C. 2008, \mnras, 385, 1242

\bibitem[{{Pfrommer} {et~al.}(2008){Pfrommer}, {En{\ss}lin}, \&
  {Springel}}]{2008MNRAS.385.1211P}
{Pfrommer}, C., {En{\ss}lin}, T.~A., \& {Springel}, V. 2008, \mnras, 385, 1211

\bibitem[{{Pinzke} \& {Pfrommer}(2010)}]{2010MNRAS.409..449P}
{Pinzke}, A. \& {Pfrommer}, C. 2010, \mnras, 409, 449

\end{thebibliography}

\end{document}